\documentclass[aps, prb, twocolumn, superscriptaddress, citeautoscript, showpacs,nofootinbib]{revtex4-2}
\usepackage{natbib}
\usepackage[dvips]{graphicx}
\usepackage[colorlinks, linkcolor = blue, citecolor = blue]{hyperref}
\usepackage{float}
\usepackage{makecell}
\usepackage{graphics}
\setcitestyle{super}
\usepackage{bm}
\usepackage{amsmath}
\usepackage{array}

\makeatletter
\newcommand*{\rom}[1]{\expandafter\@slowromancap\romannumeral #1@}
\makeatother
\DeclareUnicodeCharacter{03B1}{\ensuremath{\gamma}}

\begin{document}
\title{Analogs of Rashba-Edelstein effect from density functional theory}

\author{Karma~Tenzin}
\thanks{These authors contributed equally.}
   \affiliation{Zernike Institute for Advanced Materials, University of Groningen, Nijenborgh 4, 9747AG Groningen, Netherlands}
   \affiliation{Department of Physical Science, Sherubtse College, Royal University of Bhutan, 42007 Kanglung, Trashigang, Bhutan}
\author{Arunesh Roy}
\thanks{These authors contributed equally.}
\affiliation{Zernike Institute for Advanced Materials, University of Groningen, Nijenborgh 4, 9747AG Groningen, Netherlands}
\author{Homayoun Jafari}
   \affiliation{Zernike Institute for Advanced Materials, University of Groningen, Nijenborgh 4, 9747AG Groningen, Netherlands}
\author{Bruno Banas}
   \affiliation{Zernike Institute for Advanced Materials, University of Groningen, Nijenborgh 4, 9747AG Groningen, Netherlands}
\author{Frank T. Cerasoli}
 \affiliation{Department of Chemistry, University of California, Davis, Davis, CA 95616, USA}
\author{Mihir Date}
\thanks{Present address: Max Planck Institute of Microstructure Physics, Weinberg 2, 06114 Halle (Saale), Germany}
\affiliation{Zernike Institute for Advanced Materials, University of Groningen, Nijenborgh 4, 9747AG Groningen, Netherlands}
\author{Anooja Jayaraj}
  \affiliation{Department of Physics, University of North Texas, Denton, TX 76203, USA}
\author{Marco Buongiorno Nardelli}
 \affiliation{Department of Physics, University of North Texas, Denton, TX 76203, USA}
 \affiliation{The Santa Fe Institute, Santa Fe, NM 87501, USA}
\author{Jagoda S\l awi\'{n}ska}
\email{jagoda.slawinska@rug.nl}
 \affiliation{Zernike Institute for Advanced Materials, University of Groningen, Nijenborgh 4, 9747AG Groningen, Netherlands}
\date{\today}

\begin{abstract}
Studies of structure-property relationships in spintronics are essential for the design of materials that can fill specific roles in devices. For example, materials with low symmetry allow unconventional configurations of charge-to-spin conversion which can be used to generate efficient spin-orbit torques. Here, we explore the relationship between crystal symmetry and geometry of the Rashba-Edelstein effect (REE) that causes spin accumulation in response to an applied electric current. Based on a symmetry analysis performed for 230 crystallographic space groups, we identify classes of materials that can host conventional or collinear REE. Although transverse spin accumulation is commonly associated with the so-called 'Rashba materials', we show that the presence of specific spin texture does not easily translate to the configuration of REE. More specifically, bulk crystals may simultaneously host different types of spin-orbit fields, depending on the crystallographic point group and the symmetry of the specific $k$-vector, which, averaged over the Brillouin zone, determine the direction and magnitude of the induced spin accumulation. To explore the connection between crystal symmetry, spin texture, and the magnitude of REE, we perform first-principles calculations for representative materials with different symmetries. We believe that our results will be helpful for further computational and experimental studies, as well as the design of spintronics devices.
\end{abstract}

\maketitle

\section{Introduction}
Charge-to-spin conversion (CSC) phenomena enable all electrical creation and control of spin accumulation, \cite{manchon2015new, soumyanarayanan2016emergent} which is fundamentally important for spintronics devices such as spin-based logic \cite{manipatruni2019scalable, pham2020spin} and spin-orbit torque memories.\cite{manchon2019current, grimaldi2020single} In this context, two mechanisms arising from different microscopic origins, the spin Hall effect (SHE) and the Rashba-Edelstein effect (REE) are extensively studied in non-magnetic materials, with the aim of identifying those with optimal conversion efficiencies. In the conventional SHE, the spin accumulation ($\delta s$) occurs due to a transverse spin current ($J_S$) flowing from bulk to boundary in response to an applied charge current ($J_C$), whereby the spin polarization is perpendicular to both $J_C$ and $J_S$.\cite{sinova2015spin} In the conventional REE, a uniform spin accumulation is generated by a transverse electric current, and is caused by a non-equilibrium spin imbalance of the spin polarized bands.\cite{ganichev} Recently, unconventional SHEs allowing for arbitrary directions of electric current, spin current, and spin polarization were classified and observed,\cite{seemann2015symmetry, arunesh2021, macneill2017control, safeer2019large, ingla2022omnidirectional} expanding possibilities of electric control in spintronics devices. While unconventional REEs would allow similar functionalities, the relation between symmetry and spin texture (ST) of realistic materials and the properties of spin transport has yet to be established.

\begin{figure}
\includegraphics[width=0.48\textwidth]{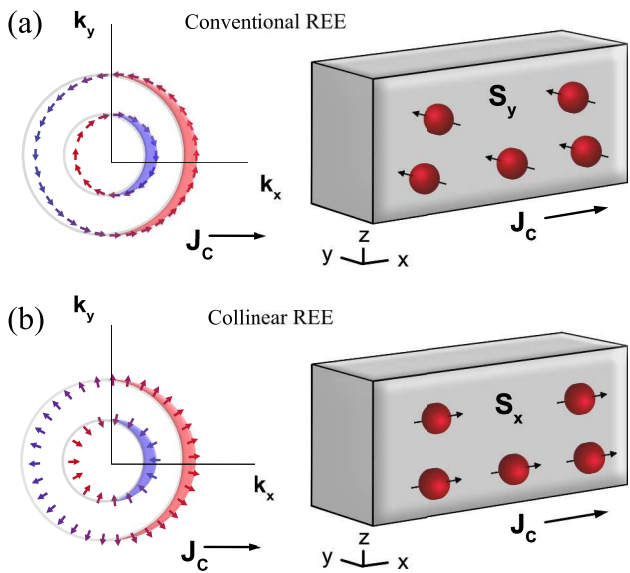}
\caption{Comparison of conventional and collinear REEs. (a) The left-hand panel shows 2D FS with purely tangential spin texture. They shift from the equilibrium in response to the electric current along $x$, to induce a net spin accumulation perpendicular to the current. Filled arcs marked in red and blue denote the generated spin imbalance. The right-hand panel shows the conventional REE in the real space. (b) The same for purely radial spin texture. In this case, the induced spin accumulation is parallel to the electric current. }
\label{fig:cisp_comparison}
\end{figure}

The conventional REE is usually associated with two-dimensional electron gas (2DEG) with circular Fermi surfaces (FS) and spin texture described by a simple Rashba model (Fig. \ref{fig:cisp_comparison}a). Analogs of REE were suggested in systems with different types of spin-orbit coupling (SOC), for example Dresselhaus (Weyl), where an induced spin accumulation can be parallel (collinear) to the electric current (see Fig. 1b). \cite{Ivchenkov1978photogalvanic, aronov1991spin, tao2021spin} However, realistic bulk materials often have complicated FS, with spin textures that deviate from models and vary across the Brillouin zone (BZ).\cite{acosta2021different} Because all \textit{k}-vectors that comprise the FS contribute to the spin accumulation, it is not straightforward to link a particular spin pattern with the occurence of conventional or collinear REE. How does a complicated spin-orbit field (SOF) in a crystal relate to the symmetry and magnitude of induced spin accumulation? To answer this question, a systematic study of the Rashba-Edelstein effect and its analogs in bulk materials is needed. While symmetry analysis can easily determine the shape of the response tensor, unveiling the role of multiple high symmetry \textit{k}-points, hosting different types of spin texture, is essential for the design of materials for efficient charge-to-spin conversion.

In this paper, we aim to link the spin texture landscape throughout the entire BZ with the presence and magnitude of specific components of the REE tensor. To this end, we first perform a systematic symmetry analysis of the response tensor for all 230 nonmagnetic space groups using the tools implemented in the Bilbao Crystallographic Server (BCS).\cite{BCS1, BCS2} Further, we use density functional theory and tight-binding Hamiltonians, generated by the post-processing code \textsc{paoflow}, to calculate both spin-orbit fields in the entire BZ and the magnitudes of the allowed REE components in representative bulk materials, illuminating potential candidates for the experimental realization of the current-induced spin accumulation with different symmetries. Based on the quantitative results, we establish the role of multiple high-symmetry $k$-points and pinpoint the mechanisms that determine conventional and collinear REE with a large magnitude.

\section{Rashba-Edelstein tensor and its symmetry}\label{sec:implementation}

\begin{figure*}
\includegraphics[width=\textwidth]{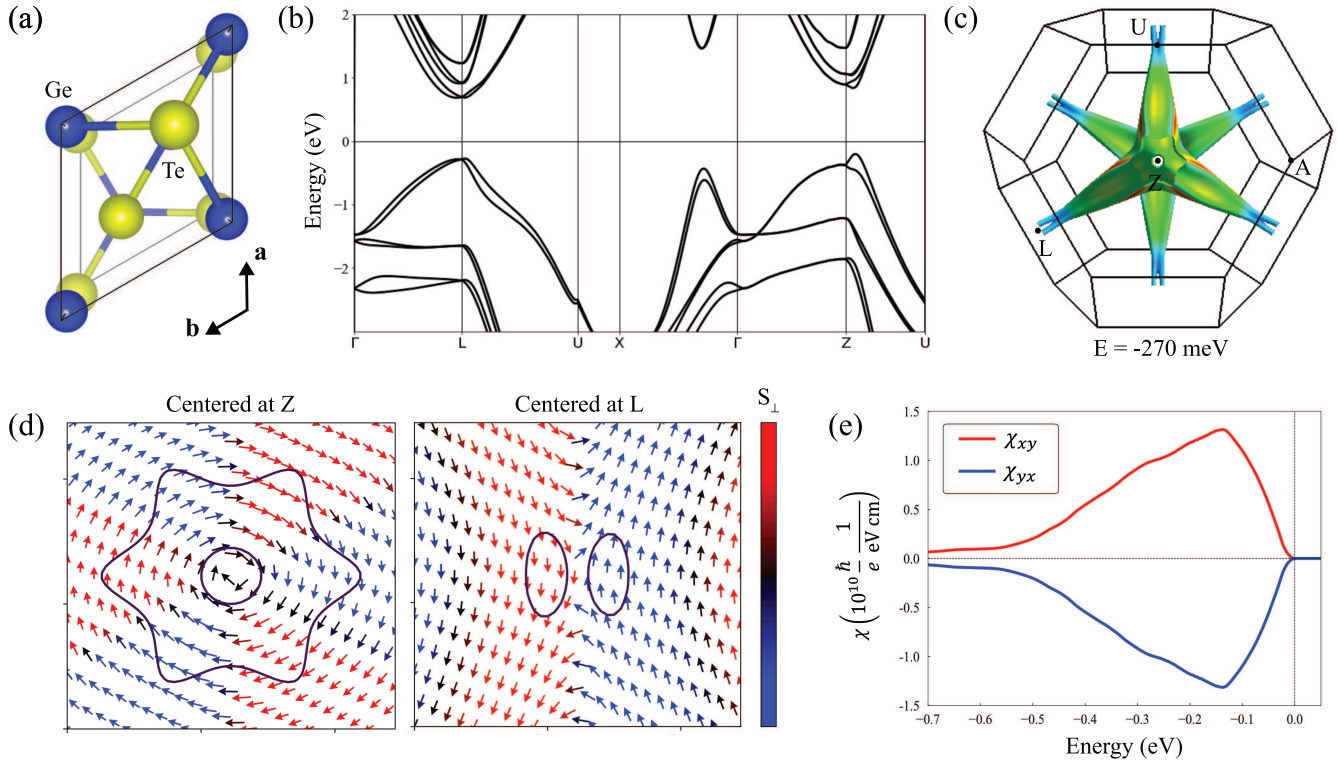}
\caption{Geometry, electronic structure and spin accumulation of ferroelectric GeTe. (a) Schematic view of the crystal structure. (b) Electronic band structure calculated along the $\Gamma-L-U-X-\Gamma-Z-U$ line. Note that the band gap is indirect and the VBM lies outside the high-symmetry line. (c) The FS at the energy $E = - 0.27$ eV below the VBM. The color scheme reflects the local band velocity.\cite{fermisurfer} (d) Left: The spin texture around the $Z$-point showing the star-shaped Fermi contour ($E = - 0.27$) with a Rashba spin texture. ($S_x$, $S_y$) components of the spin texture are represented by the arrows and the $S_z$ by the color. Right: Fermi contour at the BZ edge around the $L$ point showing a Rashba-Dresseslhaus spin texture. (e) Calculated magnitude of the $\chi$ tensor as a function of the chemical potential.}
\label{fig:gete_bs}
\end{figure*}

Let us start with the analysis of the general REE in the framework of Boltzmann transport theory. The equilibrium electronic distribution is described by the Fermi-Dirac function $f_{n\bm{k}}^{0}$ for a specific band $n$ with energy $E_{n\bm{k}}$. For a nonmagnetic material, summing expectation values of the spin operator $\bm{S}$ across all bands $n$ and momenta $\bm{k}$ in the BZ yields zero, providing no spin accumulation. When an electric field $\bm{E}$ is applied, the non-equilibrium distribution $f_{n\bm{k}} = f_{n\bm{k}}^{0} + \delta f_{n\bm{k}}$ shifts inequivalently for different bands resulting in a net spin accumulation,
\begin{equation}\label{spin_density}
\delta \bm{s} = \sum_{n\bm{k}} \langle \bm{S} \rangle_{n\bm{k}} \delta f_{n\bm{k}}.
\end{equation}
The same applied electric field changes the charge distribution in a similar way, creating a charge current density,
\begin{equation}
\bm{j}_{c} = - \frac{e}{V}\sum_{n\bm{k}} \bm{v}_{n\bm{k}} \delta f_{n\bm{k}} .
\end{equation}
where the non-equilibrium part of the distribution function is calculated as $\delta f_{n\bm{k}}^{0} = e \tau_{n\bm{k}} (\bm{E}\cdot\bm{v}_{n\bm{k}}) (\partial f_{n\bm{k}}^{0}/ \partial E_{n\bm{k}})$ with $e$ denoting the elementary charge, $\bm{v}_{n\bm{k}}$ and $E_{n\bm{k}}$ the velocity and energy of the $n^{th}$ band, and $\tau_{n\bm{k}}$ the scattering time. The current-induced spin accumulation per unit volume can be calculated as a ratio of the spin polarization $\delta \bm{s}$ and the charge current $\bm{j}_{c}$ \cite{roy2022long},
\begin{equation}\label{accmulation}
\delta \bm{s} = \chi \bm{j}^{A},
\end{equation}
where $\bm{j}^{A}$ is the applied current and
\begin{equation} \label{cisp_tensor}
\chi_{ij} = - \frac{\sum_{n\bm{k}} \langle S \rangle_{n\bm{k}}^{j} v_{n\bm{k}}^{i} (\partial f_{n\bm{k}}/\partial E_{n\bm{k}})}{e \sum_{n\bm{k}} (v_{n\bm{k}}^{i})^2 (\partial f_{n\bm{k}}/\partial E_{n\bm{k}})}
\end{equation}
defines the Rashba-Edelstein response tensor. The expectation values and velocities are computed for all band indices $n$ and momenta $\bm{k}$ in the BZ, but only the states comprising the Fermi surface will contribute to $\chi$ at zero temperature.

Several observations can be made from Eqs.(3)-(4). First, the spin accumulation induced in a solid contains contributions from the high-symmetry $k$-points as well as other $k$-points. This means that a typical model description which captures the behavior of REE originating from specific symmetry points, albeit providing analytical insights,\cite{tao2021spin} is not sufficient to characterize most realistic materials. One reason is, even those FS which are well characterized by high-symmetry points can host different spin textures at distinct \textit{k}-vectors.\cite{acosta2021different} For example, ferroelectric GeTe, a 'Rashba material', has a Rashba spin texture at the $Z$-point and a Dresselhaus-Rashba at the $L$-point, both contributing to $\chi$ (see Sec. III). Moreover, the bands, especially in metals, extend over the entire BZ, and may deviate from a well-defined ST pattern beyond the high-symmetry points. Therefore, full electronic structures calculated from first-principles are required to accurately describe the REE in bulk solids.

An arbitrary element $\chi_{ij}$ of the REE response tensor defined by Eq.(\ref{cisp_tensor}) is symmetry equivalent to $s^{ij} = \bm{v}^{i} \bm{S}^{j}$. This quantity determines whether a component will be allowed by a specific space group, namely $s^{ij}$ must be invariant under the symmetry operations (e.g. translation, improper or proper rotation, mirror reflection and inversion operation) of a space group to allow a specific element of the $\chi$ tensor \cite{arunesh2021}. In principle, the response tensor can have non-zero elements only when the space group is polar, but not all non-centrosymmetric space groups will host the REE. The same holds for optical rotation or gyration of light and in fact, the REE is determined from the crystal symmetries similarly as the gyration tensors.\cite{ganichev}

To determine the allowed form of $\chi$ for all 230 nonmagnetic space groups, we expressed the quantity $s^{ij}$, which defines the symmetry, in Jahn's notation, and used the TENSOR program available at the BCS.\cite{gallego2019automatic} Since $\chi$ (Eq.(\ref{cisp_tensor})) is axial and the two indices are independent, it can be written as $e$\{V\}\{V\} where the notation $e$ stands for the axial nature of spin, and \{\} denotes the components of a three-dimensional vector V.\cite{arunesh2021} Results of the analysis are summarized in tables for all space groups, within the Appendix. While REE is allowed in 127 space groups, we note that many possess interdependent tensor elements. For SG 1, all components of $\chi$ are allowed and are independent. As we progress toward high-symmetry space groups, more components become connected by symmetries. Tables in the Appendix will be helpful for both computational and experimental studies to explore the allowed configurations of the REE response.

The spin accumulation $\delta \bm{s}$ can be calculated from Eq.(\ref{accmulation}), using an approach based on first-principles calculations. To this end, we project \textit{ab initio} wave functions onto a set of pseudoatomic orbitals and construct accurate tight-binding (PAO) Hamiltonians \cite{agapito1, agapito2}. The PAO Hamiltonians are then interpolated to ultra-dense $k$-points meshes, a requirement for the convergence of integrated quantitites such as $\chi$. The expectation values needed to evaluate the REE tensor, such as the band velocity or spin operators are computed in \textsc{paoflow} for any specific eigenstate $\psi_{n\bm{k}}$ and followed by the evaluation of $\chi$.\renewcommand*{\thefootnote}{\fnsymbol{footnote}}\footnote{These steps are called automatically in the newly implemented routine in PAOFLOW.} We note that since the electronic structure is calculated from first-principles, we do take into account SOF at all $k$-points of the Fermi surface. Importantly, the framework of Boltzmann transport theory allows to account for the influence of the temperature ($T>0$).

\section{Analysis of representative materials}
We will now discuss in detail the properties of a few representative materials, focusing on the connection between the spin-orbit fields throughout the BZ and the symmetry and magnitude of REE. We will start with the bulk ferroelectric GeTe that simultaneously hosts Rashba and Rashba-Dresselhaus spin textures, as well as a 2D ferroelectric In$_2$Se$_3$ with purely Rashba bands close to the Fermi level. For the analysis of collinear REE, we will briefly discuss the known example of chiral Te. Last, we will show that a spin accumulation parallel to the charge current is possible also in non-chiral crystals, which will be discussed through the example of Pd$_4$Se.

\subsection*{Conventional REE in ferroelectric GeTe}
The ferroelectric Rashba semiconductor GeTe is a prototype material for ferroelectric spintronics which aims to utilize the electric polarization to control spin degrees of freedom in a non-volatile way.\cite{di2013electric, picozzi2014ferroelectric, rinaldi2018ferroelectric, prb_gete, varotto2021room} The polar GeTe crystalizes in a rocksalt structure with a rhombohedral distortion (SG 160, PG $C_{3v}$), as displayed in Fig. \ref{fig:gete_bs}a. Spontaneous polarization along the [111] direction survives up to a high temperature ($T_C$~=~700~K), \cite{rabe1987ab, kriegner2019ferroelectric} which facilitates experimental studies on the ferroelectric phase.\cite{kolobov2014ferroelectric, liebmann2016giant, krempasky2016disentangling} Figure \ref{fig:gete_bs}b shows the band structure calculated along a high-symmetry line whereas Fig. \ref{fig:gete_bs}c illustrates a FS at the energy isovalue $E=-0.27$ eV. Energy contours, in the form of a six-handed star around the $Z$-point and as smaller oval-shaped contours around the $L$-point, can be more easily seen in Fig. \ref{fig:gete_bs}d, which depicts the projections of the FS onto the BZ edges. The spin textures around these points can be interpreted in terms of the symmetry of the $k$-vectors.\cite{acosta2021different} The $Z$-point, owing to the $C_{3v}$ point group symmetry, renders the Rashba spin texture as tangential to the Fermi contours, while the $L$-point shows a more intricate pattern that can be classified, based on the $C_s$ point group symmetry, as the Rashba-Dresselhaus spin texture.\cite{acosta2021different} Although extraordinary properties of GeTe were typically associated with the Rashba spin texture, the SOF throughout the BZ cannot be entirely captured by the Rashba model.

Figure \ref{fig:gete_bs}e shows the magnitude of REE as a function of the chemical potential $E$. Since GeTe is always $p$-type doped due to the vacancies, we focused on the occupied part of the spectrum. Note that, experimentally reported values of hole concentrations ($\sim$10$^{20}$/cm$^{3}$) coincide with the energy window of around -0.2 to -0.1 eV with respect to the valence band maximum (VBM).\cite{varotto2021room} The symmetry of GeTe (SG 160) allows for only two conventional components $\chi_{xy}=-\chi_{yx}$ of the REE tensor (see Appendix). It is thus quite different from the SHE in this material, whereby multiple independent tensor elements are permitted.\cite{wang2020spin} At low energies, the spin accumulation is determined by a purely Rashba band residing around the $Z$-point. This band is strongly spin-splitted and contributes alone to the REE in a large energy window. The maximum value of 1.31 $\times 10^{10}~\frac{\hbar}{e}\frac{1}{\mathrm{eV cm}}$ at $E=-137$ meV is comparable with the values reported for other compounds,\cite{roy2022long, homayoun_wte2} and importantly, matches perfectly the energies levels accessible experimentally.\cite{varotto2021room} 
The magnitude of REE decreases at higher energies, which can be attributed to both the onset of a band with opposite spin chirality, partly compensating the spin accumulation, and the diminishing Rashba-Dresselhaus ST around the $L$-point.
We remark that the Dresselhaus component does not induce any collinear REE in this case.

\subsection*{Conventional REE in 2D-In$_2$Se$_3$}
\begin{figure*}
\includegraphics[width=\textwidth]{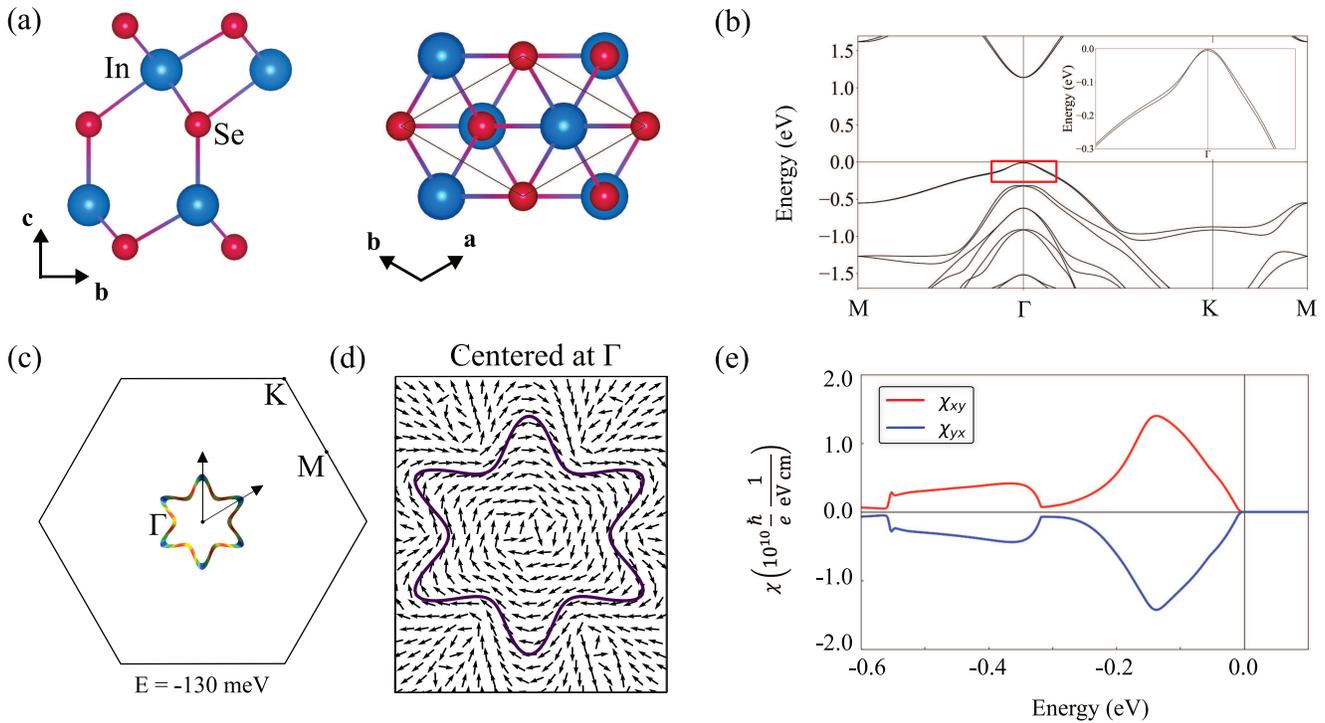}
\caption{Calculated properties of the monolayer In$_2$Se$_3$. (a) Side and top view of the crystal structure. (b) Electronic structure calculated along the high-symmetry $k$-lines. The inset shows the magnified region around the VBM. (c) Full 2D-BZ with the constant energy contour for $E=-130$ meV with respect to the VBM. (d) Zoom-in of the contour with the superimposed in-plane spin texture. ($S_x$, $S_y$) components are represented by the arrows. $S_z$ is negligible and it is omitted. (e) Calculated $\chi_{xy}$ and $\chi_{yx}$ as a function of the chemical potential. The volume equivalent was calculated using the effective thickness of In$_2$Se$_3$ equal to 10 \AA.}
\label{fig:in2se3}
\end{figure*}

Ultrathin In$_2$Se$_3$ is one of the most studied 2D materials, with different crystallographic phases observed experimentally.\cite{hu2018alpha_beta, zhou2015alpha_synthesis, xue, review_ferro} Here, we will focus only on the ferroelectric monolayer revealing both in-plane and out-of-plane electric polarization.\cite{andrew} As shown in Fig. 3a, the crystal has a thickness of five atomic layers (quintuple Se–In–Se–In–Se) and a hexagonal arrangement of atoms described by SG 156 (PG $C_{3v}$). The electronic band structure along the high-symmetry lines in the BZ, depicted in Fig 3b, shows a direct band gap at the $\Gamma$ point, $E_g = 1.13$ eV which is close to the previously reported values.\cite{andrew, ding} The FS originating from valence bands have star-shaped contours created by two states with opposite spin chiralities, as illustrated in Fig. 3c for the value of energy $E=-70$ meV.

Based on our analysis of the crystallographic point group and the wave vector point group at the $\Gamma$ point ($C_{3v}$), the SOF of the star-shaped band should be purely Rashba type,\cite{acosta2021different} which is indeed observed around the Fermi contour in Fig. 3d. The form of the $\chi$ tensor indicates that only the conventional components $\chi_{xy}=-\chi_{yx}$ can be present; their dependence on the chemical potential is plotted in Fig 3e. The magnitudes close to VBM are quite similar to GeTe, with the maximum 1.4 $\times 10^{10}~\frac{\hbar}{e}\frac{1}{\mathrm{eV cm}}$ calculated at the chemical potential of $E=-130$ meV. Such a result may seem surprising, as the spin-splitting of the topmost valence bands in GeTe is noticeably larger than in In$_2$Se$_3$ ($\sim$200 meV vs $\sim$10 meV). We believe that the large value, despite the small splitting, is related to the purely Rashba spin splitting in the considered energy window.

\subsection*{Collinear REE in chiral tellurium}
The unconventional spin accumulation induced along the direction of electric current, that we refer to as the collinear Rashba-Edelstein effect, was predicted and reported in chiral tellurium.\cite{tellurium_main, calavalle2022gate, roy2022long} Here, we will briefly recall the existing results as an important reference illustrating the relationship between the spin texture and the symmetry and magnitude of the REE. Elemental Te (SG 152 or SG 154, PG $D_3$) consists of weakly interacting helical chains running along the $z$ direction, as schematically illustrated in Fig. 4a. It is an intrinsically $p$-type doped semiconductor whose low-energy electronic structure consists of dumbbell-shaped hole pockets located at the corners of the BZ ($H$-points, see Fig. 4b). Figure 4c shows the projection of the FS onto the $k_z$ plane containing the $H$-point, which intersects the pocket at its half-length. The nearly circular Fermi contour corresponding to the energy $-30$ meV below the VBM has almost purely radial spin texture which can be described by the Weyl-type Hamiltonian. Such a pattern can be again explained via symmetry analysis; the $H$-point is characterized by the little point group $D_3$ which matches the crystallographic point group and yields a Weyl ST.\cite{acosta2021different}

The symmetry analysis of the REE tensor (see Appendix) reveals three non-zero diagonal components $\chi_{zz}$ and $\chi_{xx}=\chi_{yy}$ which originate from the radial spin texture. Figure 4d shows the tensor elements $\chi_{yy}$ and $\chi_{zz}$ calculated as a function of chemical potential $E$ with respect to the VBM. The component $\chi_{zz}$ has a maximum close to the Fermi level ($E=-$22 meV) with a magnitude of 7.99 $\times 10^{10}~\frac{\hbar}{e}\frac{1}{\mathrm{eV cm}}$, which is significantly larger than the values reported for another chiral crystal, semimetallic TaSi$_2$.\cite{roy2022long} We attribute the difference to the fact that in Te, the maximal spin accumulation is caused by only one spin-polarized band, while in a semimetal several bands with opposite chiralities mutually compensate at the Fermi level. The typical levels of $p$-type doping reported for Te samples ($10^{14}-10^{17}$/cm$^3$) allow one to access energies of up to $\sim$ -25 meV which coincides with the largest values of the REE. Overall, the results for $\chi_{zz}$ agree with the experiments,\cite{tellurium_main, calavalle2022gate, roy2022long} while the presence of the components $\chi_{xx}$ and $\chi_{yy}$ still needs to be verified via measurements.

\begin{figure*}
\includegraphics[width=\textwidth]{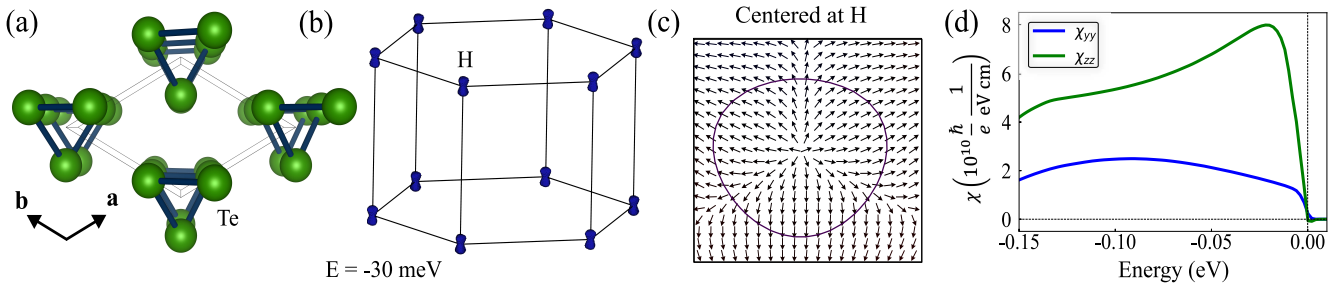}
\caption{Structure and electronic properties of the left-handed Te (SG 154). (a) Top view of the crystal structure. (b) Fermi surface at $E=-30$ meV below the VBM. (c) Projection of the pocket onto the $k_z=\pi/c$ plane around the $H$-point. The arrows represent the (S$_x$, S$_y$) components of the spin texture in the $k_x-k_y$ plane. (d) Calculated magnitude of the REE as a function of the chemical potential. The component $\chi_{xx}$ is omitted as it is equal to $\chi_{yy}$ by symmetry.}
\label{fig:te}
\end{figure*}

\subsection*{Collinear REE in non-chiral Pd$_4$Se}
Even though the presence of collinear REE is commonly associated with chirality, the symmetry analysis shows that diagonal elements of the REE tensor can be present also in non-chiral space groups, for example SG 81-82, 111-114, 121-122 (see Appendix). To find an example of such a crystal, we screened the materials database \textsc{AFLOW}\cite{aflow} and identified a non-magnetic metal with large SOC, Pd$_4$Se (SG 114, PG $D_{2d}$). It crystallizes in a structure described by the tetragonal unit cell with 10 atoms, as depicted in Fig. 5a. The Fermi surface displayed in Fig. 5c consists of several nested sheets, large ones in the center of BZ as well as cylindrical ones at the edges along the $k_z$ direction. The band structure that we plotted in Fig. 5d confirms a rather large spin-splitting of several bands. Although we calculated spin polarization of each band, the spin texture in the entire BZ is not straightforward to visualize due to the presence of multiple sheets.

The calculated current-induced spin accumulation (Fig. 5e) confirms that the diagonal components ($\chi_{xx}=-\chi_{yy}$) are non-zero. However, the estimated values around the Fermi energy are rather small, an order of magnitude lower than in the materials discussed in previous sections. To understand their origin and behavior, we analyzed spin textures of different Fermi sheets at the $E_F$. Remarkably, Weyl-type ST that gives rise to the collinear REE in Te, is not allowed at any of the high-symmetry points in Pd$_4$Se, while Weyl-Dresselhaus is possible only at the $R$ and $X$ points that do not contribute at the Fermi level. The remaining high-symmetry $k$-vectors are characterized by the point group symmetry $D_{2d}$, yielding a Dresselhaus-type spin texture. We confirm the ST by plotting the spin polarization patterns for a few Fermi contours. Fig. 5f shows the projection of the large band at the center of the BZ onto the $k_z=0$ plane while Fig. 5g and Fig. 5h illustrate the spin textures of the cylindrical bands in the respective $k_z=0$ and $k_z=\pi/c$ planes. The spin textures indeed combine tangential and radial contributions in a way that resembles the conventional Dresselhaus model.

\begin{figure*}
\includegraphics[width=\textwidth]{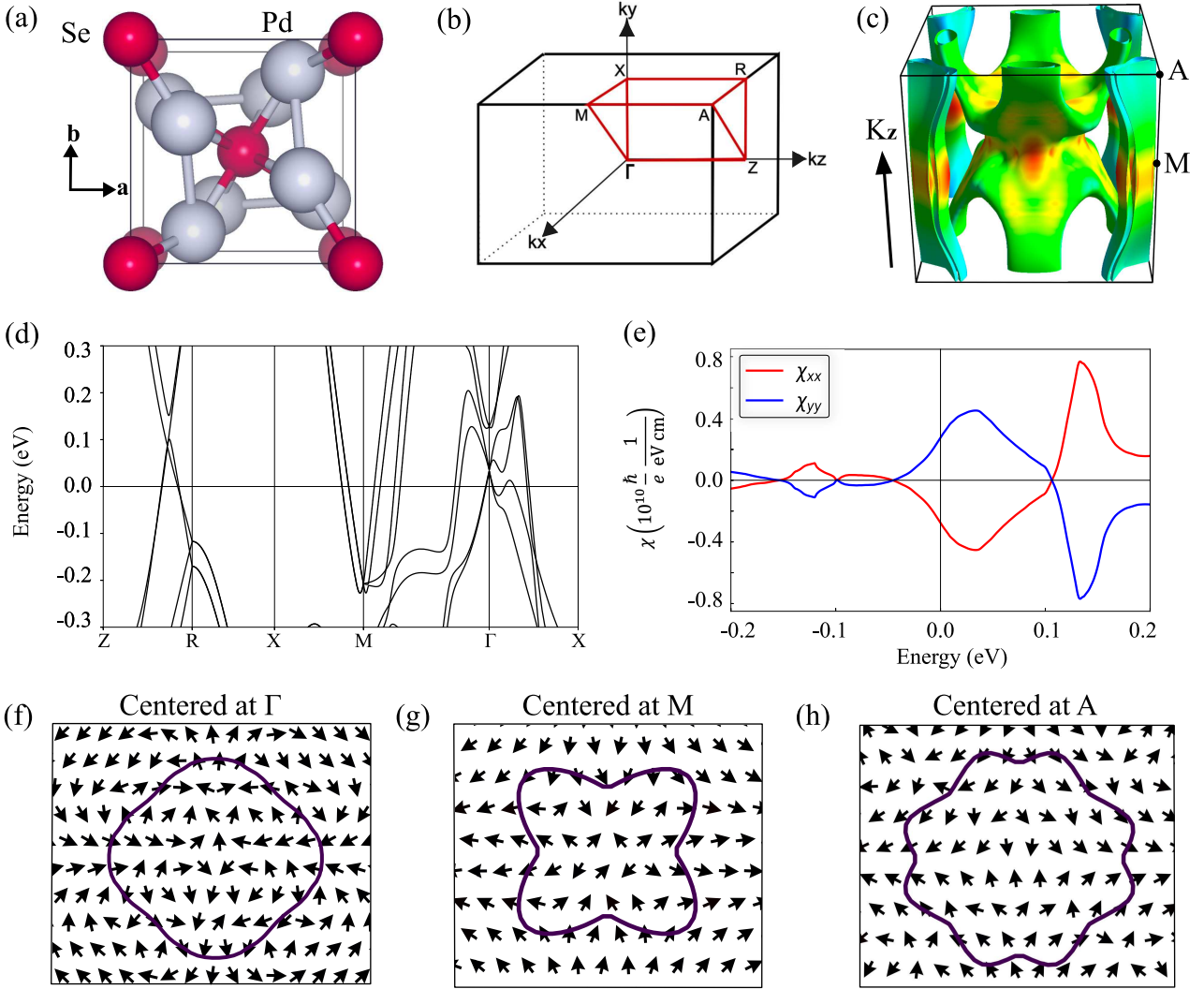}
\caption{Structural, electronic and spin transport properties of Pd$_4$Se. (a) Schematic view of the tetragonal unit cell. (b) Brillouin zone with the high-symmetry $k$-points. (c) Calculated Fermi surface ($E=E_F$) consisting of eight nested sheets. (d) Band structure calculated along the high-symmetry lines denoted in (b). (e) Calculated Rashba-Edelstein tensor elements vs. chemical potential. (f) Fermi contour ($E=E_F$) at the $k_z=0$ plane corresponding to one of the bands visible at the BZ center in (c). The arrows represent (S$_x$, S$_y$) components of the spin texture. The negligible component $S_z$ is omitted. (g) Same as (f) for one cylindrical band located along the $A-M$ line. (h) Same as (g) but plotted for the $k_z=\pi/c$ plane.}
\label{fig:pd4se}
\end{figure*}

Based on this analysis, we conclude that the radial ST component of the Dresselhaus SOF must be responsible for the collinear spin accumulation calculated at the $E_F$. Surprisingly, the tangential contributions do not yield any conventional components of the REE; further studies are therefore needed to explain when the Dresselhaus ST gives rise to the conventional and when to the collinear REE. From the results on Pd$_4$Se, we could also conjecture that the spin accumulation originating from the Dresselhaus SOC tends to be overall lower than from the Weyl; nevertheless, in case of the metallic compound with several bands of opposite spin polarizations residing around the $E_F$, the contributions to $\chi$ may just sufficiently compensate each other. A subsequent study should therefore focus on comparing several non-chiral semiconductors with the Dresselhaus ST, ideally similar to Te, to unambiguously indicate whether a pure Weyl (Rashba) ST maximizes the collinear (conventional) REE.


\section{Conclusions and perspectives}
In summary, we have performed the symmetry analysis of the REE response tensor for all 230 crystallographic space groups and explored the relationship between crystal symmetry, spin texture and properties of current-induced spin accumulation for a few representative materials. We have shown that the presence of conventional (collinear) REE cannot be easily attributed to Rashba (Weyl) SOC present in the electronic structure. We have also revealed that even a spin texture of the Dresselhaus type can cause either a conventional or unconventional REE. While spin-orbit fields accross the BZ are determined by crystallographic and wave vector point groups, which also imposes constraints on the REE tensor, some of the allowed components can be still negligible. Ultimately, the magnitudes are material dependent, as the spin accumulation density results from averaging different types of SOF patterns simultaneously contributing at the specific energy.

Because model descriptions in general suffer from the limitation of applicability, our implementation of the REE in \textsc{paoflow}, which allows estimation of spin accumulation magnitudes from first-principles methods, emerges as a very useful tool to study the effect in realistic materials. The analysis of the examples presented in this work provides an initial guidance for materials design based on high-throughput calculations. We have concluded that the occurrence of bands with single spin occupancy at a specific chemical potential would in general maximize REE. This means that, for example, doped semiconductors will be interesting candidates to explore for REE. We have also suggested that the presence of pure Rashba or Weyl spin texture may again lead to magnitudes of the REE higher than those induced by the mixture of tangential and radial spin patterns. We are overall convinced that our results will be helpful for the future search of materials with high CSC efficiency and will stimulate experimental studies of the Rashba-Edelstein effect in different configurations.

\section{Methods}
\begin{table*}[!htbp]
	\centering
	\begin{tabular}{|l|c|c|c|c|}
		\hline
		&GeTe  & In$_2$Se$_3$ &Te  & Pd$_4$Se \\
		\hline
		\thead{Lattice constants \\ (a,b,c)(\AA)}& 4.37,4.37,4.37 &  3.96,3.43,20.00&4.51,4.51,5.63  &5.23,5.23,5.64  \\
		\hline
		\thead{\textit{k}-grid\\($k_1 \times k_2 \times k_3$)}& $16 \times 16\times 16$ & $12 \times 12\times 1$&$22 \times 22\times 16$ &  $12 \times 12\times 10$  \\
		\hline
		\thead{Energy cut-off (Ry)}& 85.0 & 80.0 & 80.0 & 60.0  \\
		\hline
		\thead{$U$ parameter\\ }&  0.296 (Ge), 1.921 (Te)& \thead{3.97 (Se), 3.52 (Se),\\3.07 (Se)} & 3.81 &  0.00\\
		\hline
		\thead{nfft$_1\times$ nfft$_2\times$ nfft$_3$\\ }&$200\times 200\times 200$  &  $300\times 300\times 4$&$200\times 200\times 180$  &  $150\times 150\times 150$ \\
		\hline
	\end{tabular}
\caption{Calculation parameters that were used for each material.}
\label{detail_table}
\end{table*}
The materials calculations were performed using density functional theory (DFT) as implemented in the Quantum Espresso package \cite{giannozzi2009quantum, giannozzi2017advanced}. We treated the ion-electron interactions using the fully relativistic projector augmented wave pseudopotentials from the \textsc{pslibrary} database \cite{dal2014pseudopotentials} and we expanded the electron wave functions on plane-wave basis sets with different values of kinetic energy cutoff converged for each material. The exchange and correlation interaction was taken into account within the generalized gradient approximation (GGA) parameterized by the Perdew, Burke, and Ernzerhof (PBE) functional.\cite{pbe} The electronic structures were corrected by using a pseudo-hybrid Hubbard self-consistent approach ACBN0\cite{acbn0}. We used the Monkhorst-Pack scheme for the Brillouin zone (BZ) sampling optimized separately for each considered compound. The postprocessing calculations were performed in the \textsc{paoflow} code.\cite{paoflow1, paoflow2} The ultra-dense $k$-grids were employed to interpolate the Hamiltonians for the accurate convergence of REE. The specific calculation parameters are listed in Table \ref{detail_table}.

\newpage
\section*{Acknowledgements}
We thank Andrew Supka for the useful support. J.S. acknowledges the Rosalind Franklin Fellowship from the University of Groningen. The calculations were carried out on the Dutch national e-infrastructure with the support of SURF Cooperative (EINF-2070), on the Peregrine high-performance computing cluster of the University of Groningen and in the Texas Advanced Computing Center at the University of Texas, Austin.

\section*{Appendix} \label{sec:appendix}
To obtain the REE tensors for all crystallographic space groups, we follow the convention from the Bilbao Crystallographic Server and \textit{Physical Properties of Crystals} by Nye (Appendix B),\cite{bcs_represenation} which is also explained in detail in our previous work.\cite{arunesh2021} Below, we list the REE tensor $\chi$,
\begin{equation}
\begin{pmatrix}
     \chi_{xx} & \chi_{xy} & \chi_{xz} \\
     \chi_{yx} & \chi_{yy} & \chi_{yz} \\
     \chi_{zx} & \chi_{zy} & \chi_{zz}
\end{pmatrix}
\end{equation}
for all 230 nonmagnetic space groups, indicating independent and symmetry-related components.

\begin{table}[H]
\begin{tabular}{|m{4cm}|m{4cm}|}\hline
\begin{center}
\textbf{ \makecell{SG} \vspace{-10pt}}\end{center}  &
\makecell{$\begin{pmatrix}
     \chi_{xx} & \chi_{xy} & \chi_{xz} \\
     \chi_{yx} & \chi_{yy} & \chi_{yz} \\
     \chi_{zx} & \chi_{zy} & \chi_{zz}
\end{pmatrix}$} \\ \hline
\centering {1} &
\makecell{All independent}\\ \hline
\end{tabular}
\end{table}

\begin{table}[H]
\begin{tabular}{|m{4cm}|m{4cm}|}\hline
\begin{center}
\textbf{ \makecell{SG} \vspace{-10pt}}\end{center}  &
\makecell{$\chi = \begin{pmatrix}
     0 & 0 & 0 \\
     0 & 0 & 0 \\
     0 & 0 & 0
\end{pmatrix}$} \\ \hline
\centering {2, 10-15, 47-74, 83-88, 123-142, 147-148, 162-167, 174-176, 187-194, 200-206, 215-230} & \makecell{REE not allowed} \\ \hline
\end{tabular}
\end{table}

\begin{table}[H]
\begin{tabular}{|m{4cm}|m{4cm}|}\hline
\begin{center}
\textbf{ \makecell{SG} \vspace{-10pt}}\end{center}  &
\makecell{$\begin{pmatrix}
     \chi_{xx} & 0 & \chi_{xz} \\
     0 & \chi_{yy} & 0 \\
     \chi_{zx} & 0 & \chi_{zz}
\end{pmatrix}$} \\ \hline
\centering {3-5} &
\makecell{All independent}\\ \hline
\end{tabular}
\end{table}

\begin{table}[H]
\begin{tabular}{|m{4cm}|m{4cm}|}\hline
\begin{center}
\textbf{ \makecell{SG} \vspace{-10pt}}\end{center}  &
\makecell{$\begin{pmatrix}
     0 & \chi_{xy} & 0 \\
     \chi_{yx} & 0 & \chi_{yz} \\
     0 & \chi_{zy} & 0
\end{pmatrix}$} \\ \hline
\centering {6-9} &
\makecell{All independent}\\ \hline
\end{tabular}
\end{table}

\begin{table}[H]
\begin{tabular}{|m{4cm}|m{4cm}|}\hline
\begin{center}
\textbf{ \makecell{SG} \vspace{-10pt}}\end{center}  &
\makecell{$\begin{pmatrix}
     \chi_{xx} & 0 & 0 \\
     0 & \chi_{yy} & 0 \\
     0 & 0 & \chi_{zz}
\end{pmatrix}$} \\ \hline
\centering {16-24} &
\makecell{All $\chi_{ii}$ independent}\\ \hline
\centering {89-98, 149-155, 177-182 } &
\makecell{$\chi_{yy} = \chi_{xx}$}\\ \hline
\centering {195-199, 207-214} &
\makecell{$\chi_{zz} = \chi_{yy} = \chi_{xx}$}\\ \hline
\end{tabular}
\end{table}

\begin{table}[H]
\begin{tabular}{|m{4cm}|m{4cm}|}\hline
\begin{center}
\textbf{ \makecell{SG} \vspace{-10pt}}\end{center}  &
\makecell{$\begin{pmatrix}
     0 & \chi_{xy} & 0 \\
     \chi_{yx} & 0 & 0 \\
     0 & 0 & 0
\end{pmatrix}$} \\ \hline
\centering {25-46} &
\makecell{All $\chi_{ij}$ independent}\\ \hline
\centering {99-110, 156-161, 183-186 } &
\makecell{$\chi_{xy} = -\chi_{yx}$}\\ \hline
\centering {115-120} &
\makecell{$\chi_{xy} = \chi_{yx}$}\\ \hline
\end{tabular}
\end{table}

\begin{table}[H]
\begin{tabular}{|m{4cm}|m{4cm}|}\hline
\begin{center}
\textbf{ \makecell{SG} \vspace{-10pt}}\end{center}  &
\makecell{$\begin{pmatrix}
     \chi_{xx} & \chi_{xy} & 0 \\
     \chi_{yx} & \chi_{yy} & 0 \\
      0 & 0 & \chi_{zz}
\end{pmatrix}$} \\ \hline
\centering {75-80 ($C_4$), 143-146 ($C_3$), 168-173 ($C_6$)} &
\makecell{$\chi_{xx} = \chi_{yy}$, $\chi_{xy} = -\chi_{yx}$}\\ \hline
\end{tabular}
\end{table}

\begin{table}[H]
\begin{tabular}{|m{4cm}|m{4cm}|}\hline
\begin{center}
\textbf{ \makecell{SG} \vspace{-10pt}}\end{center}  &
\makecell{$\begin{pmatrix}
     \chi_{xx} & \chi_{xy} & 0 \\
     \chi_{yx} & \chi_{yy} & 0 \\
      0 & 0 & 0
\end{pmatrix}$} \\ \hline
\centering {81-82 ($S_4$)} &
\makecell{$\chi_{xx}= -\chi_{yy}$, $\chi_{xy} = \chi_{yx}$}\\ \hline
\end{tabular}
\end{table}

\begin{table}[H]
\begin{tabular}{|m{4cm}|m{4cm}|}\hline
\begin{center}
\textbf{ \makecell{SG} \vspace{-10pt}}\end{center}  &
\makecell{$\begin{pmatrix}
     \chi_{xx} & 0 & 0 \\
      0 & \chi_{yy} & 0 \\
      0 & 0 & 0
\end{pmatrix}$} \\ \hline
\centering {111-114, 121-122} &
\makecell{$\chi_{xx} = -\chi_{yy}$}\\ \hline
\end{tabular}
\end{table}


%

\end{document}